\documentclass[journal]{IEEEtran}
\usepackage{amsthm, amsmath, amssymb, mathtools}
\usepackage{bm}
\usepackage{hyperref}
\usepackage{orcidlink}
\usepackage{booktabs}
\usepackage{nccmath}

\newtheorem{definition}{Definition}

\newtheorem{lemma}{Lemma}

\newtheorem{proposition}{Proposition}
\newtheorem{corollary}{Corollary}

\ifCLASSINFOpdf
\else
\fi

\begin{document}

\title{Algebraic Observability for Power System Dynamic State Estimation}
\author{Akhila~Kandivalasa\,\orcidlink{0009-0009-9066-6263},~\IEEEmembership{Student~Member,~IEEE,} Marcos~Netto\,\orcidlink{0000-0001-7002-3345},~\IEEEmembership{Senior~Member,~IEEE}
\thanks{This work is supported by the National Science Foundation under Grant No. 2328241. The authors are with the Department of Electrical and Computer Engineering, New Jersey Institute of Technology, Newark, NJ 07102, USA.}
}

\maketitle

\begin{abstract}
Existing observability methods for dynamic state estimation are developed for ordinary differential equation (ODE) models, whereas power systems are naturally described by differential-algebraic equations (DAEs). Applying these methods generally requires a DAE-to-ODE transformation, usually Kron reduction, which may obscure the role of algebraic variables and network constraints. This paper presents an algebraic observability framework that operates directly on nonlinear power system DAEs and preserves their structure. A Jacobian rank test constructed from the model relations and their successive derivatives certifies the algebraic dependence of differential and algebraic variables on the known inputs and measurements. A reduced prolongation that differentiates only the stator relations keeps the number of latent indeterminates fixed as the prolongation order grows, which makes the test scalable. We establish the relationship between this rank test and hyper-regularity, an alternative framework for algebraic observability, and show that algebraic observability relates to local observability. We demonstrate algebraic observability on a single-machine decentralized setup, the WSCC 9-bus 3-machine system, and the IEEE 39-bus 10-machine system. The 39-bus test, with 257 latent indeterminates, completes in about 30 s.
\end{abstract}
\begin{IEEEkeywords}
Differential algebra, dynamic state estimation, nonlinear differential-algebraic equations, observability, power system, synchrophasor.
\end{IEEEkeywords}

\IEEEpeerreviewmaketitle

\section{Introduction}
\IEEEPARstart{M}{ost} state variables that govern power system dynamics cannot be measured directly. Stability assessment, control, and protection nonetheless depend on them \cite{Zhao2021}. Provided that the measurement configuration is adequate, dynamic state estimation (DSE) recovers these latent variables from available measurements and a system dynamic model. Whether the available measurements are sufficient to determine the states is the question of observability, and no estimator can recover a state the measurements leave undetermined. Observability analysis is therefore a prerequisite for DSE, and it remains an active research topic with open challenges \cite{Zhao2019}.

Observability analysis for DSE in power systems builds on the Hermann-Krener rank condition for nonlinear systems, which is rooted in differential geometry \cite{Hermann_Krener_1977}. The method forms an observability matrix from successive Lie derivatives of the measured outputs along the system vector field and checks its rank \cite{Rouhani2017}. The symbolic expressions resulting from Lie derivatives grow quickly and pose a severe scalability bottleneck. For the Western Systems Coordinating Council (WSCC) 3-machine, 9-bus system of \cite[\S7.3.2]{Sauer1997}, which has only 21 state variables, the computation takes about two hours \cite{Kandivalasa2026}, and it becomes intractable for larger systems.

Three approaches have been proposed to circumvent computing Lie derivatives. The empirical local Gramian \cite{Krener2009, Qi2015} replaces symbolic differentiation with numerical simulations from perturbed initial conditions, at the cost of at least two full simulations per state variable at each operating point. A derivative-free approach based on polynomial chaos \cite{Zheng2021} builds a polynomial surrogate of the map from initial states to measurements from simulations at collocation points. It then tests observability via the rank of the surrogate's coefficient matrix. The empirical local Gramian \cite{Qi2015} and derivative-free \cite{Zheng2021} methods have similar computational costs, but both still face severe scalability limitations because they require repeated numerical simulations. In sharp contrast, a graph-based approach \cite{Kandivalasa2026} reduces the computational cost to linear time in the number of nodes and edges of a graph built directly from the system's dynamic model.

Beyond computational scalability, a second, more fundamental limitation remains. The Lie derivative and all three alternative approaches above are derived explicitly for an ordinary differential equation (ODE) model. Power system dynamics, however, are represented by differential-algebraic equations (DAEs) \cite{Sauer1997, brenan1996numerical}. Differential equations describe the dynamics of machines, converters, and controllers, while algebraic equations describe the network \cite{Sauer1997}. The DAEs define a constraint manifold on which the system trajectories evolve. To apply the methods above, one first obtains an ODE model via Kron reduction \cite{Dorfler2013}, the most common DAE-to-ODE transformation for power networks. In addition to imposing a non-mild assumption that static loads behave as constant admittances, Kron reduction fundamentally alters the network structure. The well-known sparse structure of power grids becomes a fully connected graph after Kron reduction \cite{Dorfler2013}. This fundamental limitation underscores the need for an observability framework that works directly with the DAE structure of power systems.

Observability analysis for nonlinear DAEs has received limited attention in the control theory literature and almost none in power systems. Terrell \cite{terrell1997observability} builds on Hermann and Krener \cite{Hermann_Krener_1977} and constructs augmented observability Jacobians by successively differentiating a DAE model. Not surprisingly, the number of equations and the size of the resulting Jacobian grow rapidly with system size. Indeed, adding a single generator and one load bus quadruples the number of Jacobian elements \cite{dafis2003examining}, rendering the method impractical for realistic power systems.

A different route, rooted in differential algebra rather than differential geometry, has been explored far less in control theory and, to our knowledge, not at all in power systems. Rather than asking whether distinct initial conditions produce distinguishable trajectories \cite{Hermann_Krener_1977}, it asks whether algebraic relations exist between the unknown system variables and the measured quantities. This shift, from trajectories to algebraic relations, is natural for DAEs, in which the algebraic equations are an intrinsic part of the model.

This paper proposes an algebraic observability framework for power systems. It builds on the differential-algebraic theory of Diop and Fliess \cite{DiopFliess1991}. To that theory, we add a differential-polynomial formulation of the power system DAEs and a prolongation procedure that eliminates higher-order derivatives through the machine equations. A single rank test then covers the differential and algebraic variables of a nonlinear power system DAE jointly. In this theory, observability is decided by the rank of a Jacobian assembled from the model equations and a finite number of their \emph{time} derivatives. A second characterization, due to Sato \cite{sato2013algebraic}, tests algebraic observability through the hyper-regularity of a differential-operator matrix obtained from the variational equations of the DAE model. Under suitable assumptions, hyper-regularity implies local observability along regular trajectories. We show that the rank test and the hyper-regularity test answer the same question in two different mathematical languages, and that both work directly on the original power system DAE. 

We demonstrate our proposed method in a single-machine decentralized DSE setup and in a centralized setting on the WSCC 9-bus 3-machine and IEEE 39-bus 10-machine systems, all without Kron reduction. On the 39-bus system, computation takes approximately $30$~s. The contributions of this paper are as follows.

\begin{enumerate}
\item \emph{An observability test on the DAE as written.} We bring the Diop--Fliess algebraic observability framework \cite{DiopFliess1991} to power systems, where to our knowledge it has not been used, and show that a single Jacobian rank test over the model relations and finitely many of their time derivatives (Proposition~\ref{proprank}) certifies observability of the differential and algebraic variables together, with no Kron reduction and no explicit vector field.
\item \emph{A polynomial formulation and a reduced prolongation that scale.} Rotor, bus-voltage, and current-phasor angles are each replaced by a sine and cosine pair tied by a unit-circle constraint, which puts a power system model into differential-polynomial form (Table~\ref{tab:poly_dae_model}). Prolonging only the stator relations and eliminating higher machine derivatives through the machine equations keeps the latent set at 17 indeterminates per machine. On the IEEE 39-bus system, the test reaches full column rank in about 30~s, against two hours for the Lie-derivative test on the 21-state Kron-reduced WSCC model \cite{Kandivalasa2026}.
\item \emph{A bridge to local observability.} Algebraic observability in the Diop--Fliess sense is shown to be equivalent to hyper-regularity of Sato's variational differential-operator matrix \cite{sato2013algebraic} (Lemma~\ref{lem:alg_hyper_equiv}), so passing the rank test implies local observability of the nonlinear DAE along regular trajectories (Corollary~\ref{cor:diop_local_dae}).
\end{enumerate}

This paper proceeds as follows. Section \ref{obs challenges} explains the challenges of applying Lie-derivative observability analysis for power system DAEs. Section \ref{algobsframework} presents the algebraic observability framework and the rank test. Section \ref{hyper-regularity} introduces hyper-regularity and its connection to local observability. Section \ref{results} reports the results for the three test systems, and Section \ref{conclusion} concludes the paper and outlines future work.

\section{Observability Challenges in Differential-Algebraic Power System Models}\label{obs challenges}

\subsection{Scaling to electric power grids of significant size}
Consider a nonlinear dynamical system,
\begin{align}\label{eq:ode_form}
\dot{\bm{x}}&=\bm{f}(\bm{x},\bm{u}), \nonumber\\
\bm{z}&=\bm{h}(\bm{x}),
\end{align}
evolving on $\mathbb{X} \subseteq \mathbb{R}^{{n_x}}$, where the state $\bm{x} \in \mathbb{X}$, the input $\bm{u} \in \mathbb{R}^{{n_u}}$, and the output $\bm{z} \in \mathbb{R}^{{n_z}}$; $\bm{f}:\mathbb{X}\to\mathbb{X}$ defines the system dynamics, and $\bm{h}$ maps states to outputs. In this paper, by outputs, we mean measured quantities. In \eqref{eq:ode_form}, the first- and higher-order Lie derivatives of the $i$th output along the vector field $\bm{f}$ are as follows:
\begin{align}\label{eq:liederivatives}
\mathcal{L}_{\bm{f}} z_i = \frac{\partial h_i(\bm{x})}{\partial \bm{x}}\bm{f(x)}, \qquad \mathcal{L}_{\bm{f}}^{j}z_i = \frac{\partial \mathcal{L}_{\bm{f}}^{j-1}h_i(\bm{x})}{\partial \bm{x}}\bm{f(x)}.
\end{align}


\setlength{\arraycolsep}{2pt}
The observability matrix is then constructed as follows:
\begin{align}
\mathcal{O}(\bm{x}) =
\left.
\frac{\partial \mathbf{G}}{\partial \bm{x}}
\right|_{\bm{x}=\bm{x}_0},
\end{align}
with $\mathbf{G} \!=\! \left[
\begin{array}{ccccccc}
\mathcal{L}_{\bm{f}}^{0} z_1 &
\cdots &
\mathcal{L}_{\bm{f}}^{n_x-1} z_1 &
\cdots &
\mathcal{L}_{\bm{f}}^{0} z_{n_z} &
\cdots &
\mathcal{L}_{\bm{f}}^{n_x-1} z_{n_z}
\end{array}
\right]^{\mathsf{T}}$. The system is locally observable if rank$(\mathcal{O})=n_x$ \cite{Hermann_Krener_1977}. This rank test relies on the existence of an explicit vector field $\bm{f}(\bm{x},\bm{u})$ that governs the state-variable evolution. Each successive Lie derivative is computed along this vector field, so that higher-order output derivatives remain expressed purely in terms of the state variables $\bm{x}$ and the known input $\bm{u}$. The dimension of $\mathcal{O}$ is $(j+1)n_z \times n_x$, where $(j+1) \leq n_x$ \cite{Krener2009}. Even when the rank condition is satisfied at a lower order $j < n_x$, the intermediate symbolic expressions associated with the Lie derivatives rapidly expand in size.

\subsection{Extending to differential-algebraic models}
Beyond the scalability issue of symbolic expressions, the more fundamental challenge is that the Lie derivative approach is established only for ODEs, as in \eqref{eq:ode_form}. However, power system models are semi-explicit, index-1 DAEs \cite{Sauer1997}:
\begin{align}\label{eq:dae1}
\dot{\bm{x}} &= \bm{f}(\bm{x},\bm{y},\bm{u}), \nonumber\\
\bm{0} &= \bm{g}(\bm{x},\bm{y},\bm{u}), \nonumber\\
\bm{z} &= \bm{h}(\bm{x},\bm{y}), 
\end{align}
where the algebraic variables $\bm{y}\in\mathbb{R}^{n_y}$ and $\bm{g}$ encodes Kirchhoff's laws governing the network. Applying the Lie derivative construction to \eqref{eq:dae1} encounters a fundamental structural obstruction. Since the output $\bm{z}=\bm{h}(\bm{x},\bm{y})$ depends on $\bm{y}$, its derivative yields
\begin{align}
\frac{d}{dt}\bm{h} = \frac{\partial \bm{h}}{\partial \bm{x}}\bm{f}(\bm{x},\bm{y},\bm{u}) + \frac{\partial \bm{h}}{\partial \bm{y}}\dot{\bm{y}}, \label{eq:dae_lie}
\end{align}
and $\frac{\partial \bm{h}}{\partial \bm{y}}\dot{\bm{y}}$ has no counterpart in the Lie derivative framework. Indeed, \eqref{eq:dae1} does not assign $\dot{\bm{y}}$, and $\bm{g}(\bm{x},\bm{y},\bm{u})=\bm{0}$ determines $\bm{y}$ instantaneously but assigns it no dynamics. Differentiating $\bm{g}(\bm{x},\bm{y},\bm{u})=\bm{0}$, we obtain:
\begin{align}
\frac{\partial \bm{g}}{\partial \bm{y}}\dot{\bm{y}} = 
-\frac{\partial \bm{g}}{\partial \bm{x}}\bm{f}(\bm{x},\bm{y},\bm{u})
- \frac{\partial \bm{g}}{\partial \bm{u}}\dot{\bm{u}},
\label{eq:zdot}
\end{align}
which recovers $\dot{\bm{y}}$ when $\partial\bm{g}/\partial \bm{y}$ is nonsingular, as required locally for the regular semi-explicit index-1 case considered here \cite{brenan1996numerical}. However, this recovery changes the structure required for repeated output differentiation. In the original DAE formulation, the differential equations explicitly prescribe only the evolution of $\bm{x}$ through $\bm{f}(\bm{x},\bm{y},\bm{u})$, while the evolution of the algebraic variables $\bm{y}$ is specified only implicitly through the constraint $\bm{g}(\bm{x},\bm{y},\bm{u})=\bm{0}$. Hence, the complete tangent dynamics of the variable pair $(\bm{x}, \bm{y})$ are not directly available from $\bm{f}(\bm{x},\bm{y},\bm{u})$ alone. The resulting dynamics can therefore be written locally as
\begin{equation}
\begin{bmatrix}
\dot{\bm{x}}\\[1mm]
\dot{\bm{y}}
\end{bmatrix}
=
\begin{bmatrix}
\bm{f}(\bm{x},\bm{y},\bm{u})\\[1mm]
-\left(\dfrac{\partial \bm{g}}{\partial \bm{y}}\right)^{-1}
\left(
\dfrac{\partial \bm{g}}{\partial \bm{x}}\bm{f}(\bm{x},\bm{y},\bm{u})
+
\dfrac{\partial \bm{g}}{\partial \bm{u}}\dot{\bm{u}}
\right)
\end{bmatrix}. \label{eq.7}
\end{equation}
Thus, unlike an explicit ODE model, the vector field governing the complete differential--algebraic variable set is not supplied directly by the original DAE representation. Instead, it must be induced from the algebraic constraints. Moreover, successive output differentiation involves additional higher-order differential components of algebraic constraints and inputs. For example, obtaining $\ddot{\bm{y}}$ requires differentiating~\eqref{eq:zdot}, which introduces higher-order derivatives of $\bm{u}$ (e.g., $\ddot{\bm{u}}$), along with derivatives of the Jacobians, $\partial\bm{g}/\partial\bm{x}$ and $\partial\bm{g}/\partial\bm{y}$.

Consequently, the Hermann-Krener construction \cite{Hermann_Krener_1977} cannot be applied directly to \eqref{eq:dae1} by repeatedly differentiating along $\bm{f}(\bm{x},\bm{y},\bm{u})$ alone; one must first recover the missing tangent dynamics through constraint differentiation. However, this recovery comes at a structural cost: the effective vector field along which the output derivative is now expressed is no longer $\bm{f}(\bm{x},\bm{y},\bm{u})$ alone, but depends additionally on $\dot{\bm{u}}$ through \eqref{eq:zdot}. At the next differentiation order, recovering $\ddot{\bm{z}}$ requires differentiating \eqref{eq:zdot}, which introduces $\ddot{\bm{u}}$ and derivatives of the Jacobians, $\partial\bm{g}/\partial\bm{x}$ and $\partial\bm{g}/\partial\bm{y}$. At each successive order, the effective vector field changes, incorporating higher-order input derivatives and Jacobian derivatives that grow in complexity. Unlike \eqref{eq:ode_form}, where repeated Lie differentiation stays within the fixed vector field $\bm{f}(\bm{x},\bm{u})$ at every order, the DAE has no fixed vector field to differentiate along.

These limitations motivate an observability framework that works directly on the DAE structure, without transformation or variable elimination. An alternative approach, introduced by Diop and Fliess \cite{DiopFliess1991}, reformulates observability as an algebraic dependence problem. A system is observable if the state variables are algebraically dependent on measured quantities and their derivatives. The following section develops this algebraic observability framework and provides a foundation for analyzing observability in power system DAEs.

\section{Algebraic Observability}\label{algobsframework}
The algebraic observability framework departs from the conventional state-space partition of system variables into states ($\bm{x}$), algebraic variables ($\bm{y}$), inputs ($\bm{u}$), and outputs ($\bm{z}$). Instead, all system variables are classified according to their availability into two sets:
\begin{enumerate}
\item[(i)] the set of \emph{known} variables, $\bm{w}$, 
\item[(ii)] the set of \emph{latent} variables, $\bm{\ell}$, comprising all unmeasured quantities.
\end{enumerate}
In this work, the states $\bm{x}$ are unmeasured (consistent with the field practice) and included in $\bm{\ell}$. Algebraic variables $\bm{y}$, however, may belong to either set depending on their measurement availability. The known inputs $\bm{u}$ and measured outputs $\bm{z}$ are included in $\bm{w}$. Thus, the composition of $\bm{w}$ and $\bm{\ell}$ depends on the available measurements and may vary across different measurement configurations.

\begin{definition}[Ring, field, and differential field]
A ring is a set closed under addition, subtraction, and multiplication. A field is a ring in which division by any nonzero element is also allowed. A differential field is a field equipped with a differential operator satisfying the sum and product rules. \cite{kolchin1973differential}
\end{definition}

The central question of algebraic observability is whether each latent variable satisfies a finite polynomial equation whose coefficients belong to the differential field of known variables, $k\langle w\rangle$, and hence depend only on the known variables and their finite-order time derivatives \cite{DiopFliess1991}. We take $k\langle w\rangle$ to be a differential field of characteristic zero\footnote{The field $\mathbb{C}$ of complex numbers is the standard and most intuitive example of an algebraically closed field of characteristic zero, meaning every non-constant polynomial with coefficients in $\mathbb{C}$ has a root in $\mathbb{C}$, and its characteristic (the number of times $1$ must be added to itself to reach $0$) is $0$ because no finite sum equals zero.} containing the known constant parameters. The known variables $w$ generate the differential field $k\langle w\rangle$ over $k$. This field contains the known variables, their successive derivatives, and rational combinations thereof, with coefficients in $k$. Each individual expression involves only finitely many derivatives. Define $\Delta\coloneqq d/dt$. Repeated differentiation gives the operators $\Delta^j$, and $\Delta^0$ is the identity operator. The order of $\Delta^j$ is $j$. We denote the set of these operators by $\Theta=\{\Delta^j:j\geq0\}$. For a nonnegative integer $r$,
\begin{equation}
\Theta(r)=\{\Delta^0,\Delta^1,\ldots,\Delta^r\}
\end{equation}
contains the $r+1$ differentiation operators up to order $r$.

\begin{definition}[Field extension] The base field $k\langle w\rangle$, extended with the latent variables $\ell$, encompasses the unmeasured state and algebraic variables, and it gives a field extension \cite{kolchin1973differential}. Formally, a field extension of $k\langle w\rangle$ is
\begin{align}
\mathcal{E}_{k\langle w \rangle} \coloneqq k\langle w,\ell\rangle.
\end{align}
\end{definition}

The extended differential field $\mathcal{E}_{k\langle w \rangle}$ contains the elements of $w$ and $\ell$, their successive derivatives, and rational combinations thereof.

\begin{definition}[Algebraic and transcendental elements] An element $e\in \mathcal{E}_{k\langle w \rangle}$ is algebraic over $k\langle w\rangle$ if it satisfies a nonzero finite polynomial equation whose coefficients belong to $k\langle w\rangle$; that is, the coefficients are rational combinations of elements in $k\langle w \rangle$. Otherwise, $e$ is transcendental over $k\langle w\rangle$.
\end{definition}

\begin{definition}[Transcendence degree]
The transcendence degree of $\mathcal{E}_{k\langle w \rangle}$, denoted by $\operatorname{T_{deg}}\mathcal{E}_{k\langle w\rangle}$, is the maximum number of elements of $\mathcal{E}_{k\langle w \rangle}$ that are algebraically independent over $k\langle w\rangle$.
\end{definition}

Thus, if 
\begin{equation}
\operatorname{T_{deg}}\mathcal{E}_{k\langle w\rangle}=0,
\label{eq:transcendence_condition}
\end{equation}
every element of $\mathcal{E}_{k\langle w \rangle}$ is algebraic over the known field.

\begin{proposition}[Algebraic observability {\cite{DiopFliess1991}}]\label{propTrdeg}
A system represented as an extended differential field $\mathcal{E}_{k\langle w\rangle}$ is algebraically observable if and only if every latent variable in $\mathcal{E}_{k\langle w\rangle}$ is algebraic over the known differential field $k\langle w\rangle$. In particular, every latent variable satisfies a nonzero polynomial equation with coefficients in $k\langle w\rangle$.
\end{proposition}

In characteristic zero, derivatives of an algebraic element over a differential field are also algebraic over that field. Consequently, the condition in Proposition~\ref{propTrdeg} is equivalent to \eqref{eq:transcendence_condition}. Direct computation of the transcendence degree is generally inconvenient for observability analysis. The theory of K\"ahler differentials \cite{Jacobson1980} provides an alternative by relating algebraic dependence to the rank of a Jacobian. Let $k$ be a differential field. A \emph{polynomial ring} $k[\ell_1,\dots,\ell_n]$ consists of all finite sums of monomials $c\,\ell_1^{a_1}\cdots\ell_n^{a_n}$, where $c\in k$ and the exponents $a_1,\dots,a_n$ are nonnegative integers. A \emph{differential polynomial ring} $\mathcal{R}=k\langle w\rangle[\ell,\Delta\ell,\Delta^2\ell,\ldots]$ is the polynomial ring in the $\ell_i$ and their formal derivatives of all orders, each treated as an independent variable. From now on, for simplicity, we call a differential polynomial ring a \emph{ring}. We express the system equations as polynomials in the latent variables $\ell$ and their derivatives, with coefficients in the known differential field $k\langle w\rangle$. These polynomials belong to the ring
\begin{equation}
\mathcal{R}=k\langle w\rangle\{\ell\}=k\langle w\rangle[\ell,\Delta\ell,\Delta^2\ell,\ldots],
\end{equation}
$k\langle w\rangle$ is the coefficient field. The equivalent square-bracket notation explicitly treats $\ell$ and its successive derivatives $\Delta\ell,\Delta^2\ell,\ldots$ as polynomial variables. Thus, $\mathcal{R}$ contains polynomial expressions in the latent variables and their derivatives, with coefficients in the known differential field.

\begin{definition}
An \emph{ideal} is a subset of a ring closed under addition and multiplication by any element of the ring. A \emph{differential ideal} $\mathcal{I}$ is additionally closed under differentiation. 
\end{definition}
In our framework, the polynomial expressions that vanish when evaluated in $\mathcal{E}_{k\langle w\rangle}$ form a differential ideal $\mathcal{I}\subseteq\mathcal{R}$. Thus, $\mathcal{I}$ contains all polynomial relations satisfied by the system variables, including the system equations and their differential consequences.

While $\mathcal{E}_{k\langle w\rangle}$ represents the system as a differential field, its polynomial representation is described by the quotient ring
\begin{equation}
\frac{\mathcal{R}}{\mathcal{I}}=\frac{k\langle w\rangle[\ell,\Delta\ell,\Delta^2\ell,\ldots]}{\mathcal{I}}.
\end{equation}
Thus, $\mathcal{R}/\mathcal{I}$ provides a polynomial representation of the same system described by the extended differential field $\mathcal{E}_{k\langle w\rangle}$.

The coefficients of a polynomial relation establishing the algebraicity of a latent variable are rational functions of the known variables and their derivatives. These coefficients may become undefined for particular input-output trajectories if one of their denominators vanishes. Universal inputs and outputs are those satisfying the corresponding differential inequations, that is, trajectories for which the relevant denominators remain nonzero \cite{sontag1979observability, sussmann1979single}. They therefore define the generic domain on which the algebraic relations remain well defined.

For power systems, we have \eqref{eq:dae1}, whose relations originate from the differential, algebraic, and measurement equations,
\begin{equation*}
\dot{\bm{x}}-\bm{f}(\bm{x},\bm{y},\bm{u})=\bm{0},\quad
\bm{g}(\bm{x},\bm{y},\bm{u})=\bm{0},\quad
\bm{z}-\bm{h}(\bm{x},\bm{y})=\bm{0}.
\end{equation*}

In differential-polynomial form, their left-hand sides belong to $\mathcal{I}$ and provide the initial relations for the observability test. Write these equations as $p_i=0$, $i\in\{1,\ldots,n_p\}$, and collect their left-hand sides in
\begin{equation}
\mathcal{P}:=\{p_1,\ldots,p_{n_p}\}\subseteq\mathcal{I}.
\end{equation}

Since each polynomial relation $p_i=0$ holds along a system trajectory, its successive time derivatives must vanish as well. Thus, $\Delta p_i=0$, $\Delta^2 p_i=0$, and so on, wherever these derivatives exist. These are differential consequences of the original equations, rather than independently imposed relations. Since $\mathcal{I}$ is closed under differentiation, these relations belong to $\mathcal{I}$. We collect the original polynomial relations and their derivatives up to order $q$ in
\begin{equation}
\mathcal{P}^{[q]}:=\bigcup_{j=0}^{q}\{\Delta^j p_1,\ldots,\Delta^j p_{n_p}\}\subseteq\mathcal{I},
\end{equation}
which provides a finite collection of polynomial relations for constructing the observability Jacobian. 
Although these differentiated relations (often called prolongations) follow from the original equations, they may provide additional independent constraints on the latent variables and their derivatives, which are useful for assessing observability.

Let $\Theta^{[q]}(\ell)$ contain all latent variables along with their derivatives appearing in these equations, as follows:
\begin{equation}
\Theta^{[q]}(\ell):=\ell\cup\{\Delta^s\ell_i:s\geq1,\ \Delta^s\ell_i\text{ appears in }\mathcal{P}^{[q]}\}. \label{eq.15}
\end{equation}

Now, let $n_q$ denote the number of distinct elements in the collection \eqref{eq.15}. The derivative orders that appear depend on both the original equations and the number of differentiations applied.
Now the differential field extension $\mathcal{E}_{k\langle w \rangle}$ formed with the base field $k\langle w \rangle$ is represented as $\frac{k\langle w\rangle[\Theta^{[q]}(\ell)]}{\mathcal{P}^{[q]}}$. 

The theory of K\"ahler differentials \cite{Jacobson1980} connects the dimension of the differential field, its transcendence degree $\operatorname{T_{deg}}$, and the rank of the Jacobian of $\mathcal{P}^{[q]}$'s with respect to $\Theta^{[q]}(\ell)$. The dimension of the differential field is the number of independent variables $\Theta^{[q]}(\ell)$, given by
\begin{equation}
\operatorname{T_{deg}} \mathcal{E}_{k\langle w\rangle}
=
\operatorname{dim}\mathcal{E}_{k\langle w\rangle}
=
n_q - \operatorname{rank}\left(\frac{\partial\mathcal{P}^{[q]}}{\partial\Theta^{[q]}(\ell)}\right).
\label{eq:kahler_dimension}
\end{equation}

From Proposition \ref{propTrdeg}, if $\mathcal{E}_{k\langle w \rangle}$ is algebraically observable, then $\operatorname{T_{deg}}\mathcal{E}_{k\langle w \rangle}$ is zero. This demands $\operatorname{rank}\left(\frac{\partial\mathcal{P}^{[q]}}{\partial\Theta^{[q]}(\ell)}\right)=n_q$ \eqref{eq:kahler_dimension}, which can be attained if there are enough independent directions in the system of equations of $\mathcal{P}^{[q]}$ to obtain all latent variables and their derivatives in $\Theta^{[q]}(\ell)$. In simple words, the full rank implies there are no free latent variables and their derivatives $\Theta^{[q]}(\ell)$ that appear in the extended field $\mathcal{E}_{k\langle w \rangle}$. All the variables can be uniquely written as polynomials with coefficients in $k\langle w \rangle$.

\begin{proposition}\label{proprank}
Let $\mathcal{P}^{[q]}$ be the collected system relations of $\mathcal{E}_{k\langle w \rangle}$, and let $\Theta^{[q]}(\ell)$ contain all latent variables and all latent variable derivatives appearing in these relations. Then
\begin{equation}
\operatorname{rank}
\left(\frac{\partial\mathcal{P}^{[q]}}{\partial\Theta^{[q]}(\ell)}\right)=n_q,
\end{equation}
and the system $\mathcal{E}_{k\langle w \rangle}$ is algebraically observable over $k\langle w\rangle$.
\end{proposition}
Consider the following set of semi-explicit index-one DAEs:
\begin{align}
\dot{x}_1&=-x_1+y+u, &
\dot{x}_2&=x_1x_2, &
0&=x_1^2+y^2-1,
\end{align}
with known field $k\langle w \rangle$ and latent variables $\ell=\{x_1,x_2,y\}$. Suppose we have a measurement $z=h(x)$. Define
\begin{align}
p_1&=\dot{x}_1+x_1-y-u, &
p_2&=\dot{x}_2-x_1x_2,\nonumber\\
p_3&=x_1^2+y^2-1, &
p_4&=h(x)-z.
\end{align}
For a prolongation order $q$,
\begin{equation}
\mathcal P^{[q]}
=
\{\Delta^sp_i:\;i=1,\ldots,4,\;s=0,\ldots,q\},
\end{equation}
and $\Theta^{[q]}(\ell)$ contains the latent variables and all latent derivatives appearing in $\mathcal P^{[q]}$. At $q=0$,
\begin{equation}
\Theta^{[0]}(\ell)
=
\{x_1,\dot{x}_1,x_2,\dot{x}_2,y\},
\qquad n_0=5.
\end{equation}

\textit{(i) Measurement $z=x_1$:}
Here $p_4=x_1-z$. The degree-zero Jacobian is
\begin{equation}
\bm{J}_0=
\frac{\partial(p_1,p_2,p_3,p_4)}
{\partial(x_1,\dot{x}_1,x_2,\dot{x}_2,y)}
=
\begin{bmatrix}
1&1&0&0&-1\\
-x_2&0&-x_1&1&0\\
2x_1&0&0&0&2y\\
1&0&0&0&0
\end{bmatrix},
\end{equation}
which generically has $\operatorname{rank}(\bm{J}_0)\!=\!4<5$. After a prolongation, $\Theta^{[1]}(\ell)=\{x_1,\dot{x}_1,\ddot{x}_1,x_2,\dot{x}_2,\ddot{x}_2,y,\dot y\}$, $n_1=8$, and the prolonged Jacobian has $\operatorname{rank}(\bm{J}_1)=7<8$. Indeed, $z=x_1$ and $y=\dot z+z-u$, whereas
$\dot{x}_2=zx_2$ leaves one independent multiplicative degree of
freedom in $x_2$. Therefore, $n_q-\operatorname{rank}(\bm{J}_q)=1$ for every finite $q$, hence $\operatorname{T_{deg}} \mathcal{E}_{k\langle w\rangle}=1$. The system is not algebraically observable for $z=x_1$.

\textit{(ii) Measurement $z=x_2$:} Now $p_4=x_2-z$, $\operatorname{rank}(\bm{J}_0)=4<5$, and the initial relations are again insufficient. 
For the column ordering in $\Theta^{[1]}(\ell)$ above, the first prolonged Jacobian satisfies $\det(\bm{J}_1)=-4x_2y^2$, and for $x_2y\neq0$, $\operatorname{rank}(\bm{J}_1)=8=n_1$. Consequently, $n_1-\operatorname{rank}(\bm{J}_1)=0$, $\operatorname{T_{deg}} \mathcal{E}_{k\langle w\rangle}=0$, and Proposition~\ref{proprank} establishes algebraic observability.

This example therefore illustrates the complete rank-test procedure: the original and prolonged polynomial relations determine $\Theta^{[q]}(\ell)$, and algebraic observability is attained when the Jacobian reaches full column rank.

\section{Algebraic Observability and Hyper-Regularity}
\label{hyper-regularity}
An alternative algebraic observability test is based on hyper-regularity, a property of matrices over skew polynomials. The nonlinear DAE model \eqref{eq:dae1} is said algebraically observable if a polynomial matrix derived from its DAE representation is hyper-regular \cite{sato2013algebraic}. In this section, we connect the Jacobian rank criterion introduced in section~\ref{algobsframework} to the hyper-regularity condition. This connection provides a bridge between the algebraic observability rank test and local observability \cite{Hermann_Krener_1977} for the class of index-one DAE systems considered here. In particular, the connection holds along a regular and feasible trajectory, where the polynomial matrices and transformations involved in the hyper-regularity analysis remain well-defined. Along such a trajectory, hyper-regularity implies complete observability of the corresponding linearized system and, consequently, local observability of the original nonlinear DAE. 

Consider the DAE system \eqref{eq:dae1}. A meromorphic function is analytic on its domain except at isolated poles, and therefore excludes essential singularities. Let $\mathcal{M}$ denote the field of meromorphic functions generated by the system variables and a finite number of derivatives of the known inputs,
$\{x_i,y_j,u_k^{(m)} \!\mid\! 1 \le\! i \le {n_x},\; 1 \!\le j \le n_y,\; 1 \le k \le n_u,\; m \ge 0\},$ and let $\bm{f},\,\bm{g},\,\bm{h}\in\mathcal{M}$. The system dynamics equip $\mathcal{M}$ with the time-derivative operator $\Delta$. Hence, for any admissible function $\phi\in\mathcal{M}$, its time derivative remains in the same field, i.e., $\Delta\phi=d\phi/dt\in\mathcal{M}$. Then we introduce the infinitesimal variations $d\bm{x}$, $d\bm{y}$, and $d\bm{u}$, which represent arbitrarily small changes in the states, algebraic variables, and inputs, respectively. We extend the time-derivation operator $\Delta$ to these variations, allowing both the system quantities and their infinitesimal changes to be treated under the same differential structure. Finite-order differential operators with coefficients in $\mathcal{M}$ then form the ring $\mathcal{D}:=\mathcal{M}[\Delta]$.

Because $\Delta$ differentiates not only the system variables but also the meromorphic coefficients multiplying them, the order of multiplication matters; thus, $\mathcal{D}$ is a noncommutative skew-polynomial differential-operator ring. The polynomial matrix used in the hyper-regularity formulation is constructed over this ring. This construction is consistent with the differential-field setting introduced in Section~\ref{algobsframework}. In the rank test formulation, the system is described over a differential field closed under differentiation and with universal inputs and outputs, while Sato \cite{sato2013algebraic} uses the meromorphic field $\mathcal{M}$ together with the differential operator $\Delta$. Therefore, although the two approaches use different algebraic machinery, both formulate observability on the same type of differential-function ground: the admissible system functions and their derivatives remain within a well-defined differential field, and the analysis excludes singular points where the required expressions are undefined.

We now return to \eqref{eq:dae1}, where $d\bm{x}$, $d\bm{y}$, $d\bm{u}$, and $d\bm{z}$ denote infinitesimal variations between neighboring system trajectories. These variations should be distinguished from the time-derivation operator $\Delta$: for example, $d\bm{x}$ represents a small variation in $\bm{x}$, whereas $\Delta(d\bm{x})$ represents the time derivative of that variation. Taking the first-order variation of the differential, algebraic, and measurement equations in \eqref{eq:dae1} gives
\begin{align}
    \Delta(d\bm{x})-\bm{f}_{\bm{x}}d\bm{x}-\bm{f}_{\bm{y}}d\bm{y} &= \bm{f}_{\bm{u}}d\bm{u}, \\
    \bm{g}_{\bm{x}}d\bm{x}+\bm{g}_{\bm{y}}d\bm{y} &= -\bm{g}_{\bm{u}}d\bm{u}, \\
    \bm{h}_{\bm{x}}d\bm{x}+\bm{h}_{\bm{y}}d\bm{y} &= d\bm{z},
\end{align}
where the subscripts denote the corresponding partial derivatives. Collecting these relations yields
\begin{equation}\label{eqal}
\mathbf{A_L}
\begin{bmatrix}d\bm{x}\\d\bm{y}\end{bmatrix}
=
\mathbf{A_R}
\begin{bmatrix}d\bm{u}\\d\bm{z}\end{bmatrix}
\end{equation}
with
\begin{equation}
\mathbf{A_L}
=
\begin{bmatrix}
\Delta\bm{I}-\bm{f}_{\bm{x}} & -\bm{f}_{\bm{y}} \\
\bm{g}_{\bm{x}} & \bm{g}_{\bm{y}} \\
\bm{h}_{\bm{x}} & \bm{h}_{\bm{y}}
\end{bmatrix},
\qquad
\mathbf{A_R}
=
\begin{bmatrix}
\bm{f}_{\bm{u}} & 0 \\
-\bm{g}_{\bm{u}} & 0 \\
0 & \bm{I}
\end{bmatrix}.
\end{equation}
Since the entries of these matrices are meromorphic functions and finite-order differential operators in $\Delta$, $\mathbf{A_L}$ and $\mathbf{A_R}$ are defined over the differential-operator ring $\mathcal{D}$. Hence, \eqref{eq:dae1} is algebraically observable if the matrix $\mathbf{A_L}$ is hyper-regular.

\begin{definition}
A matrix $\mathbf{A_L}$ is hyper-regular if there exist unimodular matrices $\bm{U}$ and $\bm{V}$ over $\mathcal{D}$ such that
\begin{equation}
    \bm{U}\mathbf{A_L}\bm{V}
    =
    \left[\begin{array}{cc}
       \bm{I}_{n_x+n_y}  &  \bm{0}
    \end{array}\right]^{\mathsf{T}}.
    \label{UPV}
\end{equation}
A matrix is unimodular if it is invertible over $\mathcal{D}$ and its inverse also belongs to $\mathcal{D}$. Therefore, $\bm{U} \bm{U}^{-1}= \bm{I}$, $\bm{V} \bm{V}^{-1}= \bm{I}$, and $\bm{U}, \bm{V} \in \mathcal{D}$.
\end{definition}

Equation~\eqref{UPV} represents the simplest Jacobson form of $\mathbf{A_L}$ over the noncommutative differential-operator ring $\mathcal{D}$. This implies that the no-differential-operator factor remains in its normal form. Collecting the latent variables in a vector $\bm{\ell} = [\bm{x}^\mathsf{T}\ \bm{y}^\mathsf{T}]^{\mathsf{T}}$ and the known variables in $\bm{w} = [\bm{u}^\mathsf{T}\ \bm{z}^\mathsf{T}]^{\mathsf{T}}$, \eqref{eqal} becomes,
\begin{equation}\label{UPV2}
    \mathbf{A_L}\,d\bm{\ell}
    =
    \mathbf{A_R}\,d\bm{w}.
\end{equation}
Since $\mathbf{A_L}$ is hyper-regular, \eqref{UPV2} implies:
\begin{enumerate}
\item Hyper-regularity removes any independent differential degree of freedom associated with the latent variables. In other words, their infinitesimal variations are completely constrained by the known variables and a finite number of their derivatives.
\item Consequently, $d\bm{\ell}$ can be expressed in terms of known variables $d\bm{w}$ and its coefficients without additional derivation operators left. 
\end{enumerate}
The Diop--Fliess \cite{DiopFliess1991} and Sato \cite{sato2013algebraic} formulations express the same loss of independent differential freedom in the latent variables through two different algebraic representations. In the differential-field formulation, this is characterized by algebraicity of $\ell$ over $k\langle w\rangle$, or equivalently by zero transcendence degree. In the differential-operator formulation, it is characterized by hyper-regularity of $\mathbf{A_L}$.

\begin{lemma}\label{lem:alg_hyper_equiv}
For a regular index-one DAE system \eqref{eq:dae1}, under the common differential-field assumptions introduced above,
\begin{equation}
\ell \ \text{is algebraic over }k\langle w\rangle
\ \Longleftrightarrow\
\mathbf{A_L}\ \text{is hyper-regular}.
\end{equation}
\end{lemma}

\begin{proof}
Suppose that $\mathbf{A_L}$ is hyper-regular. Then $\mathbf{L}\!\coloneqq\!\bm{V}[\bm{I}\;\;\bm{0}]\bm{U}$ satisfies $\mathbf{L}\mathbf{A_L}=\bm{I}$, and \eqref{UPV2} yields $d\bm{\ell}=\mathbf{L}\mathbf{A_R}d\bm{w}$. Hence, when $d\bm{w}=0$, no nonzero independent differential variation of $\ell$ remains. The relative differential module therefore has zero rank; equivalently, the extension has zero transcendence degree over $k\langle w\rangle$. Thus, $\ell$ is algebraic over $k\langle w\rangle$. Conversely, suppose that $\ell$ is algebraic over $k\langle w\rangle$. Then the corresponding extension has zero transcendence degree, so the relative differential module contains no independent differential direction in $d\bm{\ell}$ when $d\bm{w}=\bm{0}$. For the regular index-one representation considered here, this is equivalent to $\mathbf{A_L}$ having full column rank over the differential-operator ring, and hence $\mathbf{A_L}$ is hyper-regular.
\end{proof}

Lemma~\ref{lem:alg_hyper_equiv} connects the algebraic-dependence condition used in Section~\ref{algobsframework} with Sato's differential-operator characterization. The importance of this lemma is that hyper-regularity also provides a connection to local observability \cite{Hermann_Krener_1977}. Along a regular trajectory, algebraic observability in the hyper-regularity sense implies complete observability of the corresponding linearized system, which in turn implies local observability of the original nonlinear DAE \cite{sato2013algebraic}. A regular trajectory is a feasible analytic trajectory $(\bm{x}^{*}(t),\bm{y}^{*}(t),\bm{u}^{*}(t))$ for which the coefficient functions entering $\mathbf{A_L}$, $\mathbf{A_R}$, and the unimodular transformations remain well defined.

To apply this result to the Diop--Fliess rank criterion, we restrict attention to the universal input--output trajectories introduced in Section~\ref{algobsframework}, for which the algebraic relations of the latent variables remain well defined. If such a specialization $w^{*}(t)$ generates a feasible analytic trajectory for which the differential-operator matrices $\mathbf{A_L}$, $\mathbf{A_R}$ and the corresponding unimodular transformations are well defined, then the trajectory is regular in the sense required by \cite{sato2013algebraic}. This connects the algebraic-observability result from the Jacobian rank test to local observability via Lemma~\ref{lem:alg_hyper_equiv}.

\begin{corollary}\label{cor:diop_local_dae}
Consider the index-one nonlinear DAE system \eqref{eq:dae1}. If the Diop--Fliess Jacobian rank criterion of Proposition~\ref{proprank} is satisfied, then \eqref{eq:dae1} is locally observable along any universal input--output specialization.
\end{corollary}

\begin{proof}
By Proposition~\ref{proprank}, satisfying the Diop--Fliess Jacobian rank criterion implies that the latent variables $\ell$ are algebraic over the known differential field $k\langle w\rangle$. Lemma~\ref{lem:alg_hyper_equiv} then gives the equivalent hyper-regularity of $\mathbf{A_L}$. Hence, for any feasible analytic trajectory satisfying the stated universal input--output condition and Sato's regularity requirements, the corresponding linearized system is completely observable by \cite{sato2013algebraic}. Complete observability of this linearization implies local observability of the original nonlinear DAE along that trajectory.
\end{proof}

\section{Power-System Implementation and Results}
\label{results}
We implement both algebraic observability frameworks on power system DAEs\footnote{Code available: \url{https://github.com/akk2797/algebraicobservabilityforpsdse}} using the fourth-order generator model equipped with an IEEE Type-1 exciter; see (7.1)--(7.7), (7.13)--(7.14), and (7.32)--(7.35) in \cite{Sauer1997}. Hereafter, $i$ indexes generator buses, $j$ indexes non-generator buses, $k$ indexes all buses in network summations, and $b$ denotes a generic bus of either type. For a generator connected at bus $i$, let the terminal-voltage and terminal-current phasors be $\bar V_i = V_i e^{j\theta_i}$, $\bar I_i = I_i e^{j\phi_i}$. The corresponding voltage and current components in the rotor $dq$
reference frame are
\begin{align}
V_{d_i}
&=
V_i\sin(\delta_i-\theta_i),
&
V_{q_i}
&=
V_i\cos(\delta_i-\theta_i),
\\
I_{d_i}
&=
I_i\sin(\delta_i-\phi_i),
&
I_{q_i}
&=
I_i\cos(\delta_i-\phi_i).
\label{eq:dq_phasor_conversion}
\end{align}
Hence, the rotor-frame variables $V_{d_i}$, $V_{q_i}$, $I_{d_i}$, and
$I_{q_i}$ can be eliminated.
For example, the $q$-axis transient-voltage equation (7.1) in \cite{Sauer1997} is $T_{do_i}\dot E'_{q_i}=E_{fd_i}-E'_{q_i}-(X_{d_i}-X'_{d_i})I_{d_i}$, and it becomes,
\begin{align*}
T_{do_i}\dot E'_{q_i}
-
E_{fd_i}
+
E'_{q_i}
+
(X_{d_i}-X'_{d_i})
I_i\sin(\delta_i-\phi_i)
=
0.
\end{align*}
The remaining equations are expressed in the same manner. Although this representation eliminates the rotor-frame voltage and current components, it is not yet suitable for the rank test introduced in Section~\ref{algobsframework}. The Diop--Fliess rank test requires a differential-polynomial representation of the system. This creates a difficulty when the latent rotor angle $\delta_i$ appears inside trigonometric functions, since $\sin\delta_i$ and $\cos\delta_i$ are not polynomial functions of the indeterminate $\delta_i$. Therefore, rather than retaining $\delta_i$ as a latent indeterminate, we replace the rotor-angle coordinate by the algebraic variables $S_{\delta_i}:=\sin\delta_i$ and $C_{\delta_i}:=\cos\delta_i$. Their dependence is enforced through the polynomial constraint $S_{\delta_i}^2+C_{\delta_i}^2-1=0$ and the corresponding differential relations. Using $\dot{\delta}_i=\omega_i-\omega_s$, the rotor-angle dynamics are equivalently represented as
\begin{align}
\dot{S}_{\delta_i}-C_{\delta_i}(\omega_i-\omega_s)&=0, &
\dot{C}_{\delta_i}+S_{\delta_i}(\omega_i-\omega_s)&=0.
\end{align}
Thus, $\delta_i$ need not be retained as a separate latent indeterminate. 
We apply the same polynomial transformation to the bus-voltage angles $\theta_b$ and generator current-phasor angles $\phi_i$. Accordingly, we define
\begin{equation*}
S_{\theta_b}\coloneqq\sin\theta_b,\; C_{\theta_b}\coloneqq\cos\theta_b,\;
S_{\phi_i}\coloneqq\sin\phi_i,\; C_{\phi_i}\coloneqq\cos\phi_i.
\end{equation*}

When these angles are measured, their sine and cosine components are treated as known quantities in the coefficient field $k\langle w\rangle$. Otherwise, they are retained as latent indeterminates subject to the corresponding unit-circle constraints. Using the trigonometric identities,
\begin{align}
\sin(\delta_i-\phi_i)&=S_{\delta_i}C_{\phi_i}-C_{\delta_i}S_{\phi_i}, \notag\\
\cos(\delta_i-\phi_i)&=C_{\delta_i}C_{\phi_i}+S_{\delta_i}S_{\phi_i}, \notag\\
\sin(\delta_i-\theta_i)&=S_{\delta_i}C_{\theta_i}-C_{\delta_i}S_{\theta_i}, \notag\\
\cos(\delta_i-\theta_i)&=C_{\delta_i}C_{\theta_i}+S_{\delta_i}S_{\theta_i},
\end{align}
the differential and algebraic equations can be expressed as polynomials in the latent indeterminates. 

For the network equations, the voltage-angle differences involve $\theta_b$, $\theta_k$, and the known admittance angle $\alpha_{bk}$. Applying the same trigonometric identities, we define
\begin{align}
\mathcal{C}_{bk}&\!=\!(C_{\theta_b}C_{\theta_k}\!+S_{\theta_b}S_{\theta_k})\cos\alpha_{bk}\!+\!(S_{\theta_b}C_{\theta_k}\!-C_{\theta_b}S_{\theta_k})\sin\alpha_{bk},\\
\mathcal{S}_{bk}&\!=\!(S_{\theta_b}C_{\theta_k}\!-C_{\theta_b}S_{\theta_k})\cos\alpha_{bk}\!-\!(C_{\theta_b}C_{\theta_k}\!+S_{\theta_b}S_{\theta_k})\sin\alpha_{bk},
\end{align}
which represent $\cos(\theta_b-\theta_k-\alpha_{bk})$ and $\sin(\theta_b-\theta_k-\alpha_{bk})$, respectively. Since the network parameters are known, these expressions are polynomial in the lifted voltage-angle variables. Similarly, the generator current-injection terms are expressed using $C_{\theta_i}C_{\phi_i}+S_{\theta_i}S_{\phi_i}$ and $S_{\theta_i}C_{\phi_i}-C_{\theta_i}S_{\phi_i}$. Thus, we represent the generator and network equations, together with the ZIP load model, in differential-polynomial form. Table~\ref{tab:poly_dae_model} in the Appendix shows the resulting model.

\subsection*{Case 1: Decentralized DSE}
In this case, we analyze observability for a single generator, which is a decentralized DSE configuration. We first apply the rank test and then examine the hyper-regularity of the associated matrix $\mathbf{A_L}$. For the subsequent multi-machine case studies, we apply only the rank test, as computing unimodular transformations and the Jacobson form becomes computationally demanding as the model size increases.

\subsubsection*{1A: Rank test extension to power system models}
For the decentralized DSE setup, the terminal voltage and current phasors of machine $i$ are locally available, and $V_{\mathrm{ref},i}$ is a known input. Because we do not model the turbine-governor system, we also treat $T_{M_i}$ as a known input. If one models the turbine-governor system, $T_{M_i}$ becomes a latent variable. Accordingly, $w=\left\{V_i,\theta_i,I_i,\phi_i,T_{M_i},V_{\mathrm{ref},i}\right\}$, and the differential field $k\langle w\rangle$ contains these quantities and their required time derivatives. 
Since $\theta_i$ and $\phi_i$ are measured, their sine and cosine components can be computed from the measurements. For the polynomial rank test, the known differential field is enlarged to include $S_{\theta_i}$, $C_{\theta_i}$, $S_{\phi_i}$, and $C_{\phi_i}$, together with their required derivatives. The machine parameters are treated as known constants. The latent indeterminates are selected from the unmeasured quantities that remain after the polynomial transformation, together with the additional variables introduced by that transformation. Although $\delta_i$ itself is eliminated, its derivative $\dot{\delta}_i$ remains in the polynomial relations through $\dot{\delta}_i-\omega_i+\omega_s=0$, and also appears in the differential identities associated with $S_{\delta_i}$ and $C_{\delta_i}$. Therefore, $\dot{\delta}_i$ is retained as a latent indeterminate in the present formulation. The latent-variable set used to construct the Jacobian is
\begin{align}\label{eq.37}
\Theta(\ell)=\big\{&
\omega_i,E'_{q_i},E'_{d_i},E_{fd_i},V_{R_i},R_{F_i},
\dot{\delta}_i,\dot{\omega}_i,\dot{E}'_{q_i},\dot{E}'_{d_i},
\nonumber\\
&\dot{E}_{fd_i},\dot{V}_{R_i},\dot{R}_{F_i},
S_{\delta_i},C_{\delta_i},
\dot{S}_{\delta_i},\dot{C}_{\delta_i}
\big\},
\end{align}
so that $n_q= |\Theta(\ell)|=17$. 
The initial polynomial set $\mathcal{P}^{[0]}$ contains the seven original machine differential relations, the two stator algebraic equations, and the three auxiliary relations
\begin{align}
S_{\delta_i}^2+C_{\delta_i}^2-1=0,\;
\dot{S}_{\delta_i}-C_{\delta_i}\dot{\delta}_i=0,\;
\dot{C}_{\delta_i}+S_{\delta_i}\dot{\delta}_i=0.
\end{align}
Hence, $|\mathcal{P}^{[0]}|=12$. The corresponding Jacobian is
\begin{equation}
\bm{J}^{[0]}=
\frac{\partial\mathcal{P}^{[0]}}
{\partial\Theta(\ell)}, \qquad \operatorname{rank}\bm{J}^{[0]}=12<17.
\end{equation}

\begin{figure}[!t]
\centering
\includegraphics[width=1.0\linewidth]{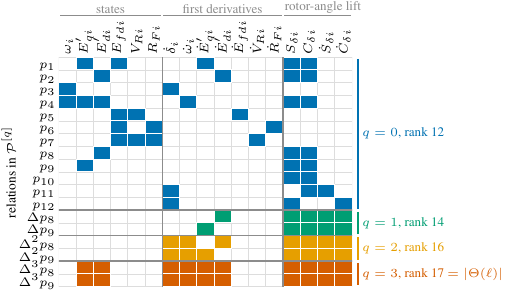}
\caption{Evolution of the single-machine observability Jacobian with successive prolongations. Rows represent the original and differentiated polynomial relations, while columns represent the 17 latent indeterminates. Colored entries indicate nonzero Jacobian elements at each prolongation order. The rank increases from 12 to 14, 16, and 17, attaining full column rank at $q=3$.}
\label{fig.JacStruct}
\end{figure}

\begin{figure}[!t]
\centering
\includegraphics[width=1.0\linewidth]{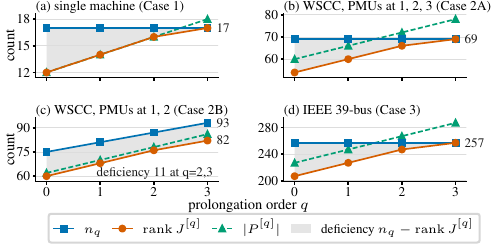}
\caption{Evolution of the observability Jacobian rank with successive prolongations for (a) the decentralized single-machine system, (b) WSCC 9-bus system with PMUs at buses 1, 2, and 3, (c) WSCC 9-bus system with PMUs at buses 1 and 2, and (d) IEEE 39-bus system. The plots compare the number of latent indeterminates $n_q$, polynomial relations $|\mathcal{P}^{[q]}|$, and Jacobian rank $\operatorname{rank}J^{[q]}$. Full column rank is attained in Cases 1, 2A, and 3, whereas Case 2B retains a rank deficiency of 11 through the tested prolongation orders.}
\label{fig.rank_progression}
\end{figure}

To obtain additional polynomial relations, the two stator algebraic equations are successively differentiated. At each prolongation order, higher-order machine derivatives are eliminated using the machine differential equations. The derivatives of $S_{\delta_i}$ and $C_{\delta_i}$ are handled using their differential identities, while derivatives of measured quantities remain in the known differential field. This reduced prolongation procedure keeps the Jacobian columns fixed at the same 17 latent indeterminates. 
Fig.~\ref{fig.JacStruct} illustrates the construction of the observability Jacobian from the initial polynomial relations and their successive prolongations.
Fig. \ref{fig.rank_progression}(a) summarizes the rank evolution. The Jacobian rank increases from $12$ at the initial level to $14$ and $16$ after the first and second prolongations, respectively. At the third prolongation, $\operatorname{rank}\bm{J}^{[3]}=17=|\Theta(\ell)|$, and the Jacobian therefore attains full column rank. Hence, the decentralized single-machine model satisfies the Diop--Fliess algebraic-observability rank criterion. It takes less than $1$~s to compute prolongations and the rank of the corresponding Jacobian test in MATLAB, as shown in Fig. \ref{fig:CompTimes}.

\subsubsection*{1B: Hyper-Regularity of the power system model}
We examine hyper-regularity using both the original seven-state meromorphic DAE and the polynomial representation in Table \ref{tab:poly_dae_model}. We use the original formulation here to present the test, since Sato's differential-operator framework permits meromorphic coefficients and therefore does not require the trigonometric terms to be polynomialized. The same test was also repeated on the lifted $(S_{\delta_i},C_{\delta_i})$ representation and led to the same hyper-regularity conclusion, providing an independent consistency check between the two representations.

The algebraic variables $\{V_i,\, \theta_i,\, I_i,\, \phi_i\}$ are available in the decentralized case. We take $\bm{u}_i=[V_i\ \theta_i\ T_{M_i}\ V_{\mathrm{ref},i}]^{\mathsf T}$ and $\bm{z}_i=[I_i\ \phi_i]^{\mathsf T}$. We express the DAE system \eqref{eq:dae1} in the form
\begin{equation}
\mathbf{A_L}_i
\begin{bmatrix}d\bm{x}_i\\d\bm{y}_i\end{bmatrix}
=
\mathbf{A_R}_i
\begin{bmatrix}d\bm{u}_i\\d\bm{z}_i\end{bmatrix},
\end{equation}
with
\begin{equation*}
\mathbf{A_L}_i=
\begin{bmatrix}
\Delta\bm{I}_7-\bm{f}_{i,\bm{x}} & -\bm{f}_{i,\bm{y}}\\
\bm{g}_{i,\bm{x}} & \bm{g}_{i,\bm{y}}\\
\bm{0}_{2\times7} & \bm{H}_{i,\bm{y}}
\end{bmatrix},
\qquad
\mathbf{A_R}_i=
\begin{bmatrix}
\bm{f}_{i,\bm{u}} & \bm{0}_{7\times2}\\
-\bm{g}_{i,\bm{u}} & \bm{0}_{2\times2}\\
\bm{0}_{2\times4} & \bm{I}_2
\end{bmatrix},
\end{equation*}
where $\bm{H}_{i,\bm{y}}=[\bm{0}_{2\times2}\ \bm{I}_2]$ and $\Delta a=a\Delta+\dot a$ for every trajectory-dependent coefficient $a$. Here, $\bm{g}_i$ includes two stator algebraic equations. The first block row of $\mathbf{A_L}_i$ contains seven generator differential equations ($7\times11$), the second contains two algebraic relations ($2\times11$), and the third contains two measurement equations ($2\times11$). Thus, the matrix has $7+2+2=11$ rows, while its $7+4=11$ columns correspond to the seven differential states $d\bm{x}_i$ and four algebraic variables $d\bm{y}_i$. Therefore, $\mathbf{A_L}_i$ is a $11\times11$ differential-operator matrix. We computed its Jacobson form in Mathematica using the noncommutative operator rule, and obtained
\begin{equation}
\bm{U}_i\mathbf{A_L}_i\bm{V}_i=
\begin{bmatrix}\bm{I}_{11}\\\bm{0}_{2\times11}\end{bmatrix},
\end{equation}
for unimodular matrices $\bm{U}_i$ and $\bm{V}_i$. Hence, all eleven invariant diagonal entries are units and $\mathbf{A_L}_i$ is hyper-regular.

We repeated the computation for the differential-polynomial model and obtained the same hyper-regularity conclusion. The decentralized single-machine system is therefore algebraically observable under both criteria considered in this work: the Diop--Fliess rank test attains full column rank after finite prolongation, and the associated variational differential-operator matrix satisfies the hyper-regularity condition. These two independent checks confirm algebraic observability of the local machine model for the given measurement configuration.

\subsection*{Case 2: Centralized DSE, WSCC 9-bus 3-machine System}
We next extend the algebraic observability rank test to the WSCC three-machine nine-bus system in a centralized DSE setting. The system consists of three synchronous generators connected through a nine-bus network. The original DAE model comprises 21 generator differential equations, six stator algebraic equations, and 18 network equations, for a total of 45 equations. To obtain the differential-polynomial formulation, each generator introduces three auxiliary rotor-angle relations. These add nine relations for the three generators, increasing the total to 54 equations.

\subsubsection*{2A: Measurements collected from all three machine buses}
In this measurement configuration, voltage and current phasors are collected at buses 1--3. Mechanical torques and voltage references are treated as known inputs. Since the voltage angles at buses 4--9 are unmeasured, we introduce six bus-voltage angle constraints, resulting in 60 initial polynomial relations. The known quantities are collected as
\begin{equation}
w=
\{V_i,\theta_i,I_i,\phi_i,T_{M_i},V_{\mathrm{ref},i}\}_{i=1}^{3}.
\end{equation}

\begin{figure}[!t]
\centering
\includegraphics[width=1.0\linewidth]{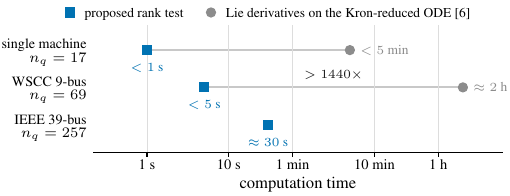}
\caption{Computational scalability of the proposed algebraic observability rank test for nonlinear power system DAE model, compared with the conventional Lie-derivative approach on Kron-reduced ODE models.}
\label{fig:CompTimes}
\end{figure}

The latent-variable set contains the same 17 machine indeterminates defined in Case~1 for each generator, together with the unmeasured network variables at buses $4$--$9$. 
Since the non-generator-bus voltage angles are latent, they are represented through their sine and cosine variables rather than retained directly. Hence,
\begin{equation}
\Theta(\ell)=
\bigcup_{i=1}^{3}\Theta_i(\ell)
\;\cup\;
\bigcup_{j=4}^{9}
\{V_j,S_{\theta_j},C_{\theta_j}\},
\end{equation}
where $\Theta_i(\ell)$ denotes the 17-variable single-machine latent set in \eqref{eq.37}. Therefore, $n_q= |\Theta(\ell)|=3(17)+6(3)=69$. For each unmeasured network angle, we include the additional algebraic relation $S_{\theta_j}^{2}+C_{\theta_j}^{2}-1=0$, $j\in\{4,\ldots,9\}$ in the initial polynomial set. We form the initial polynomial set $\mathcal{P}^{[0]}$ from the transformed generator and network relations together with the rotor- and network-angle algebraic constraints. To generate additional independent relations, only the two stator constraints of each generator are successively differentiated. We recursively eliminate the higher-order machine derivatives introduced during this process using the machine differential equations and the lifted rotor-angle relations. Since the terminal phasors of all three generators are measured, their derivatives remain in $k\langle w\rangle$, and the latent-variable set remains fixed at 69 variables for all prolongation orders. At each order, $\bm{J}^{[q]}=\partial \mathcal{P}^{[q]} / \partial \Theta(\ell)$ is evaluated, and its rank is compared with the 69 latent-variable columns. Fig. \ref{fig.rank_progression}(b) summarizes the resulting rank evolution. The rank increases from $54$ at the initial level to $60$ and $66$ after the first and second prolongations, respectively. At the third prolongation, $\operatorname{rank}\bm{J}^{[3]}=69=|\Theta(\ell)|$, so the Jacobian attains full column rank. Hence, the WSCC three-machine nine-bus system satisfies the algebraic-observability rank criterion for this measurement configuration. The complete rank test, up to $q=3$ with $n_q=69$, required less than $5$ s, compared with approximately two hours reported for the Lie-derivative test on the same system's Kron-reduced ODE model \cite{Kandivalasa2026} (Fig. \ref{fig:CompTimes}).

\subsubsection*{2B: Measurements collected from only two machine buses}
We next remove the terminal voltage and current phasor measurements at generator bus 3 while retaining those at buses 1 and 2. 
The known quantities are therefore
\begin{equation}
w=\{V_i,\theta_i,I_i,\phi_i\}_{i=1}^{2}
\cup\{T_{M_i},V_{\mathrm{ref},i}\}_{i=1}^{3}.
\end{equation}
Consequently, the terminal quantities associated with machine 3 are no longer elements of the known differential field and must be included among the latent variables. 
At degree zero, the removal of measurements at bus 3 introduces six additional latent indeterminates, $\{V_3,I_3,S_{\theta_3},C_{\theta_3},S_{\phi_3},C_{\phi_3}\}$, increasing the total from 69 to 75. Since both $\theta_3$ and $\phi_3$ are unmeasured, their corresponding angle constraints are also included, increasing the initial polynomial relation count from 60 to 62. Unlike the fully measured case, prolongation of the third generator's stator constraints introduces derivatives of the unmeasured terminal quantities. These derivatives cannot be absorbed into $k\langle w\rangle$ and must therefore be appended to the latent-variable set at each differentiation order. As a result, the number of unknown columns increases with $q$. At each prolongation order, the six stator algebraic equations and the two angle constraints corresponding to the unmeasured angles $\theta_3$ and $\phi_3$ are differentiated, introducing eight additional polynomial relations. Consequently, the number of polynomial relations increases from 62 to 86, while the number of latent indeterminates increases from 75 to 93 over $q=0,1,2,3$. Fig. \ref{fig.rank_progression}(c) summarizes the corresponding rank results. The unknown count increases from $75$ to $93$ over the tested prolongations, while the Jacobian rank remains strictly below the number of latent-variable columns. Full column rank is therefore not attained through degree three or beyond, and this measurement configuration does not satisfy the algebraic-observability rank criterion.

\subsection*{Case 3: Extension to the IEEE 39-Bus 10-Machine System}
To examine the scalability of the proposed rank-test formulation, we extend the analysis to the IEEE 39-bus 10-machine New England system. We obtain the network topology, branch parameters, shunt data, and nominal loads from MATPOWER \texttt{case39} and retain them as fixed coefficients in the differential-polynomial model. No Kron reduction is performed. Terminal voltage and current phasor measurements are assumed available at all ten generator buses, while the mechanical torques and voltage references are treated as known inputs. The latent variables and their required derivatives are retained symbolically during the construction of the observability Jacobian. The ten generators contribute $10(17)=170$ latent indeterminates, while the 29 non-generator buses contribute three variables each, $(V_j,S_{\theta_j},C_{\theta_j})$, yielding an additional $29(3)=87$ variables. Hence, $n_q=|\Theta(\ell)|=170+87=257$. Fig.~\ref{fig.rank_progression}(d) summarizes the rank-test results. The Jacobian rank increases with successive prolongations and attains full column rank at $q=3$, $\operatorname{rank}\bm{J}^{[3]}=257=n_q$. Hence, the IEEE 39-bus system satisfies the algebraic-observability rank criterion under the considered measurement configuration. The computation required $\approx30$~s as shown in Fig. \ref{fig:CompTimes}. This computational efficiency is attributed to the reduced prolongation procedure, in which only the 20 stator algebraic equations are successively differentiated, while higher-order machine derivatives are eliminated using the governing differential equations. Consequently, the number of latent indeterminates remains fixed at 257, and the number of polynomial relations increases by only 20 at each prolongation order. Furthermore, the sparse structure of the machine and network equations helps limit the computational burden associated with Jacobian construction and rank evaluation. These results demonstrate the computational tractability of the proposed formulation for the IEEE 39-bus system.

\section{Conclusion and Future Work}\label{conclusion}
This paper developed an algebraic-observability framework for power system nonlinear DAEs that preserves the generator, network, and static load equations without reducing them to ODEs. The proposed method uses a differential-polynomial Jacobian rank test to assess whether the latent variables are algebraic over the differential field generated by the known inputs and measurements, while hyper-regularity provides a complementary characterization and connects algebraic observability with local observability. We demonstrated the formulation on a single-machine decentralized DSE setup, the WSCC 9-bus 3-machine system, and the IEEE 39-bus 10-machine system. Full column rank was attained after finite prolongation for the single-machine case, for the WSCC system with measurements at all three generator buses, and for the IEEE 39-bus system. When we removed measurements at one WSCC generator bus, the latent-variable set increased under prolongation, and full column rank was not recovered within the tested orders, illustrating how measurement availability affects algebraic observability. Future work will consider quantitative observability metrics, and measurement-placement design. We will also explore graph-theoretic structure to further reduce the computational burden for larger systems.

\appendix\label{app}
Table~\ref{tab:poly_dae_model} presents the power system model in differential-polynomial form.

\begin{table}[!t]
\centering
\caption{Differential-polynomial DAE model used for algebraic-observability analysis}
\label{tab:poly_dae_model}

\begin{minipage}{0.98\columnwidth}

\textit{Generator differential equations:}
\begin{align*}
T_{do_i}\dot{E}'_{q_i}
-E_{fd_i}
+E'_{q_i}
+(X_{d_i}-X'_{d_i})I_i
\left(S_{\delta_i}C_{\phi_i}-C_{\delta_i}S_{\phi_i}\right)
&=0,\\
T_{qo_i}\dot{E}'_{d_i}
+E'_{d_i}
-(X_{q_i}-X'_{q_i})I_i
\left(C_{\delta_i}C_{\phi_i}+S_{\delta_i}S_{\phi_i}\right)
&=0,\\
\dot{\delta}_i-\omega_i+\omega_s&=0,\\
\dot S_{\delta_i}
-C_{\delta_i}(\omega_i-\omega_s)
&=0,\\
\dot C_{\delta_i}
+S_{\delta_i}(\omega_i-\omega_s)
&=0,\\
\frac{2H_i}{\omega_s}\dot{\omega}_i
-T_{M_i}+D_i(\omega_i-\omega_s)&\\
+E'_{d_i}I_i
\left(S_{\delta_i}C_{\phi_i}-C_{\delta_i}S_{\phi_i}\right)
+E'_{q_i}I_i
\left(C_{\delta_i}C_{\phi_i}+S_{\delta_i}S_{\phi_i}\right)&\\
+(X'_{q_i}-X'_{d_i})I_i^2
\left(S_{\delta_i}C_{\phi_i}-C_{\delta_i}S_{\phi_i}\right)
\left(C_{\delta_i}C_{\phi_i}+S_{\delta_i}S_{\phi_i}\right)
&=0,\\
T_{E_i}\dot{E}_{fd_i}
+(K_{E_i}+S_{E_i})E_{fd_i}
-V_{R_i}
&=0,\\
T_{F_i}\dot{R}_{F_i}
+R_{F_i}
-\frac{K_{F_i}}{T_{F_i}}E_{fd_i}
&=0,\\
T_{A_i}\dot{V}_{R_i}
+V_{R_i}
-K_{A_i}R_{F_i}
+\frac{K_{A_i}K_{F_i}}{T_{F_i}}E_{fd_i}
-K_{A_i}(V_{\mathrm{ref},i}-V_i)
&=0.
\end{align*}

\begin{fleqn}[0pt]
\begin{equation*}
\textit{Rotor-angle algebraic constraint:}
\qquad\qquad\qquad\,\,
S_{\delta_i}^{2}+C_{\delta_i}^{2}-1=0.
\end{equation*}
\end{fleqn}

\begin{fleqn}[0pt]
\begin{equation*}
\textit{Current-phasor angle algebraic constraint:}
\quad\;\;\;\,
S_{\phi_i}^{2}+C_{\phi_i}^{2}-1=0.
\end{equation*}
\end{fleqn}

\begin{fleqn}[0pt]
\begin{equation*}
\textit{Bus-voltage angle algebraic constraint:}
\qquad\quad\;\;\,
S_{\theta_b}^{2}+C_{\theta_b}^{2}-1=0.
\end{equation*}
\end{fleqn}

\textit{Stator algebraic equations:}
\begin{align*}
E'_{d_i}
-V_i\left(S_{\delta_i}C_{\theta_i}-C_{\delta_i}S_{\theta_i}\right)
-R_{s_i}I_i
\left(S_{\delta_i}C_{\phi_i}-C_{\delta_i}S_{\phi_i}\right)&\\
+X'_{q_i}I_i
\left(C_{\delta_i}C_{\phi_i}+S_{\delta_i}S_{\phi_i}\right)
&=0,\\
E'_{q_i}
-V_i\left(C_{\delta_i}C_{\theta_i}+S_{\delta_i}S_{\theta_i}\right)
-R_{s_i}I_i
\left(C_{\delta_i}C_{\phi_i}+S_{\delta_i}S_{\phi_i}\right)&\\
-X'_{d_i}I_i
\left(S_{\delta_i}C_{\phi_i}-C_{\delta_i}S_{\phi_i}\right)
&=0.
\end{align*}

\textit{Generator-bus network equations:}
\begin{align*}
V_iI_i(C_{\theta_i}C_{\phi_i}+S_{\theta_i}S_{\phi_i})
+P_{L_i}(V_i)
-\sum_{k=1}^{n}V_iV_kY_{ik}\mathcal{C}_{ik}
&=0,\\
V_iI_i(S_{\theta_i}C_{\phi_i}-C_{\theta_i}S_{\phi_i})
+Q_{L_i}(V_i)
-\sum_{k=1}^{n}V_iV_kY_{ik}\mathcal{S}_{ik}
&=0.
\end{align*}

\textit{Non-generator-bus network equations:}
\begin{align*}
P_{L_j}(V_j)
-\sum_{k=1}^{n}V_jV_kY_{jk}\mathcal{C}_{jk}
&=0,\\
Q_{L_j}(V_j)
-\sum_{k=1}^{n}V_jV_kY_{jk}\mathcal{S}_{jk}
&=0.
\end{align*}

\textit{ZIP load model:}
\begin{align*}
P_{L_b}(V_b)
&=P_{L_b}^{0}\!\left[
a_{P_b}\!\left(\frac{V_b}{V_b^{0}}\right)^2
+b_{P_b}\!\left(\frac{V_b}{V_b^{0}}\right)
+c_{P_b}\right],\\
Q_{L_b}(V_b)
&=Q_{L_b}^{0}\!\left[
a_{Q_b}\!\left(\frac{V_b}{V_b^{0}}\right)^2
+b_{Q_b}\!\left(\frac{V_b}{V_b^{0}}\right)
+c_{Q_b}\right].
\end{align*}

\end{minipage}
\end{table}

\bibliographystyle{IEEEtran}

\begin{thebibliography}{10}
\providecommand{\url}[1]{#1}
\csname url@samestyle\endcsname
\providecommand{\newblock}{\relax}
\providecommand{\bibinfo}[2]{#2}
\providecommand{\BIBentrySTDinterwordspacing}{\spaceskip=0pt\relax}
\providecommand{\BIBentryALTinterwordstretchfactor}{4}
\providecommand{\BIBentryALTinterwordspacing}{\spaceskip=\fontdimen2\font plus
\BIBentryALTinterwordstretchfactor\fontdimen3\font minus \fontdimen4\font\relax}
\providecommand{\BIBforeignlanguage}[2]{{%
\expandafter\ifx\csname l@#1\endcsname\relax
\typeout{** WARNING: IEEEtran.bst: No hyphenation pattern has been}%
\typeout{** loaded for the language `#1'. Using the pattern for}%
\typeout{** the default language instead.}%
\else
\language=\csname l@#1\endcsname
\fi
#2}}
\providecommand{\BIBdecl}{\relax}
\BIBdecl

\bibitem{Zhao2021}
J.~Zhao \emph{et~al.}, ``Roles of dynamic state estimation in power system modeling, monitoring and operation,'' \emph{IEEE Trans. Power Syst.}, vol.~36, no.~3, pp. 2462--2472, 2021.

\bibitem{Zhao2019}
------, ``Power system dynamic state estimation: Motivations, definitions, methodologies, and future work,'' \emph{IEEE Trans. Power Syst.}, vol.~34, no.~4, pp. 3188--3198, 2019.

\bibitem{Hermann_Krener_1977}
R.~Hermann and A.~Krener, ``Nonlinear controllability and observability,'' \emph{IEEE Trans. Autom. Control}, vol.~22, no.~5, pp. 728--740, 1977.

\bibitem{Rouhani2017}
A.~Rouhani and A.~Abur, ``Observability analysis for dynamic state estimation of synchronous machines,'' \emph{IEEE Trans. Power Syst.}, vol.~32, no.~4, pp. 3168--3175, 2017.

\bibitem{Sauer1997}
P.~W. Sauer and M.~A. Pai, \emph{Power System Dynamics and Stability}.\hskip 1em plus 0.5em minus 0.4em\relax Prentice Hall, 1997.

\bibitem{Kandivalasa2026}
A.~Kandivalasa and M.~Netto, ``Graph approach for observability analysis in power system dynamic state estimation,'' in \emph{IEEE PESGM}, 2026, pp. 1--5.

\bibitem{Krener2009}
A.~J. Krener and K.~Ide, ``Measures of unobservability,'' in \emph{IEEE CDC}, 2009, pp. 6401--6406.

\bibitem{Qi2015}
J.~Qi, K.~Sun, and W.~Kang, ``Optimal {PMU} placement for power system dynamic state estimation by using empirical observability {G}ramian,'' \emph{IEEE Trans. Power Syst.}, vol.~30, no.~4, pp. 2041--2054, 2015.

\bibitem{Zheng2021}
Z.~Zheng \emph{et~al.}, ``Observability analysis of a power system stochastic dynamic model using a derivative-free approach,'' \emph{IEEE Trans. Power Syst.}, vol.~36, no.~6, pp. 5834--5845, 2021.

\bibitem{brenan1996numerical}
K.~E. Brenan, S.~L. Campbell, and L.~R. Petzold, \emph{Numerical Solution of Initial-Value Problems in Differential-Algebraic Equations}.\hskip 1em plus 0.5em minus 0.4em\relax SIAM, 1995.

\bibitem{Dorfler2013}
F.~Dorfler and F.~Bullo, ``Kron reduction of graphs with applications to electrical networks,'' \emph{IEEE Trans. Circuits Syst. I: Regul. Pap.}, vol.~60, no.~1, pp. 150--163, 2013.

\bibitem{terrell1997observability}
W.~J. Terrell, ``Observability of nonlinear differential algebraic systems,'' \emph{Circuits Syst. Signal Process.}, vol.~16, no.~2, pp. 271--285, 1997.

\bibitem{dafis2003examining}
C.~J. Dafis and C.~Nwankpa, ``Examining characteristics of an observability formulation for nonlinear power systems,'' in \emph{IEEE ISCAS}, vol.~3, 2003, pp. 395--398.

\bibitem{DiopFliess1991}
S.~Diop and M.~Fliess, ``On nonlinear observability,'' in \emph{ECC}, vol.~1, Grenoble, France, Jul. 1991, pp. 152--157.

\bibitem{sato2013algebraic}
K.~Sato, ``Algebraic observability of nonlinear differential algebraic systems with geometric index one,'' in \emph{IEEE CDC}, 2013, pp. 2582--2587.

\bibitem{kolchin1973differential}
E.~R. Kolchin, \emph{Differential algebra \& algebraic groups}.\hskip 1em plus 0.5em minus 0.4em\relax Academic press, 1973, vol.~54.

\bibitem{Jacobson1980}
N.~Jacobson, \emph{Basic Algebra II}.\hskip 1em plus 0.5em minus 0.4em\relax San Francisco, CA: W. H. Freeman and Company, 1980.

\bibitem{sontag1979observability}
E.~D. Sontag, ``On the observability of polynomial systems, i: Finite-time problems,'' \emph{SIAM J. Control Optim.}, vol.~17, no.~1, pp. 139--151, 1979.

\bibitem{sussmann1979single}
H.~J. Sussmann, ``Single-input observability of continuous-time systems,'' \emph{Math. Syst. Theory}, vol.~12, pp. 371--393, 1979.

\end{thebibliography}

\end{document}